\documentclass[3p,preprint]{elsarticle} 
\usepackage{epsfig}
\usepackage{amsfonts}
\usepackage{amsmath}
\usepackage{amssymb}
\usepackage{xcolor}
\usepackage[normalem]{ulem}

\usepackage[titletoc,toc,title]{appendix}

\usepackage[bookmarks=true]{hyperref}

\begin{document}

\title{Excitonic states in spherical layered quantum dots}

\author[famaf]{Mariano Garagiola}
\ead{mgaragiola@famaf.unc.edu.ar}

\author[famaf]{Omar Osenda\corref{corresp}}
\ead{osenda@famaf.unc.edu.ar}


\cortext[corresp]{Corresponding author}

\address[famaf]{Facultad de Matem\'atica, Astronom\'{\i}a, F\'{\i}sica y 
Computaci\'{o}n, Universidad Nacional de C\'ordoba and Instituto de 
F\'{i}sica Enrique Gaviola, CONICET, Ciudad Universitaria, C\'ordoba, 
Argentina X5000HUA}

\begin{abstract}
The properties of excitons formed in spherical quantum dots are studied using 
the $\mathbf{k}\cdot\mathbf{p}$ method within the Hartree approximation. The 
spherical quantum dots considered have a central core and  several concentric 
layers of different semiconductor materials that are modeled as a succession of 
potential wells and barriers. The $\mathbf{k}\cdot\mathbf{p}$ Hamiltonian and 
the Coulomb equations for the electron-hole pair are solved using a 
self-consistent iterative method. The calculation of the spectrum of the empty 
quantum dot and the electron-hole pair is performed by means of a very accurate 
numerical approximation. It is found that the exciton binding energy as a 
function of the core radius of the quantum dot shows a strong non-linear 
behaviour. In particular, for quantum dots with two potential wells, the binding 
energy presents a large steep change. This last behaviour is explained in terms 
of the polarization charges at the interfaces between different materials and 
the matching conditions  for the eigenfunctions. 

\end{abstract}

\date{\today}

\maketitle

\section{Introduction}\label{sec:introduccion}

The methods to produce multi-layered spherical quantum dots are well 
established for some years now 
\cite{kortan1990,zhou1993,mews1994,kouwenhoven2001}. Their two-dimensional 
counterpart, also known 
as quantum rings, have a similar development. Both kind of settings, two and 
three-dimensional, have been  used to experimentally test the foundations of 
the Quantum Mechanics theory 
and an increasing number of applications. 

The confinement of electrons in semiconductor quantum dots is owed to different 
mechanisms. In assembled quantum dots, the confinement is produced by the 
mismatch  between the energy bands of the different materials that compose the 
heterostructure. This mechanism is so used that the possible semiconductor 
heterojunctions are classified by its band alignment, types one, two and three, 
which are also known as straddling, staggered and broken gaps, respectively.

Quantum dot quantum well structures (QDQW), {\em i.e.} the spherical structures 
that have a spherical core made up of a wide gap semiconductor, a first layer 
surrounding the core (also known as the well) that is made up of a narrow gap 
material , 
and a second layer made up with a  wide gap material, have been extensively 
studied, in particular the  CdS/HgS/CdS QDQW 
\cite{haus1993,schooss1994,bryant1995,banyai1992,chang1998}. 

The number of layers to consider in a given structure is chosen to meet some 
design requirement. In the case QDQW structures, the desired effect was the 
increasing of the cross section of the confined electrons, with respect to a 
simple QD with a similar total radius \cite{muller2004}.

The energy gaps characteristic of the semiconductor materials that made up the 
most common multi-layered quantum  dots are bounded between 1.5 and 3 eV or 
even higher values so, at low or even ambient temperatures, the conduction band 
is empty and the promotion of electrons from the valence band to the conduction 
one is achieved applying electromagnetic fields with  optical or near optical 
frequencies. As a matter of fact, the study of this type of structures and 
materials is 
fueled by its possible application to construct solar cells or 
light emitting diodes \cite{kang2016,shen2013}. Once an electron 
is promoted from the valence band to the conduction one the hole left by the 
electron interacts with it, forming an electron-hole pair. more commonly known 
as an exciton \cite{yoffe1993,woggon1995}. Of course, this is a simple, and 
partial, description of the 
many-body response of all the electrons involved 
\cite{wang1994,franceschetti1997,ougut1997}. Anyway, 
the absorption lines observed in many experiments confirm that the description 
in terms of two interacting particles of opposite charge has a very broad 
range of validity and the theoretical calculations made in models formulated 
within this description predict the experimental findings more or less 
accurately .

The simplest model that captures the two-particle description of an exciton is 
formulated using the Effective Mass Approximation (EMA) 
approach \cite{schooss1994,bryant1995,chang1998,franceschetti1997,billaud2009,
yao2010} . In this 
approximation the electron and hole physics is described using 
Schr\"odinger-like equations that depend on effective parameters for the masses 
of both electron and hole. Besides, the interaction between the pair is given 
by a Coulomb potential that includes a macroscopic dielectric constant, and the 
confinement potential is provided by piecewise potentials that model the band 
structure profile of the nanostructure. The model with all the ingredients 
mentioned before is also known as the one-band EMA, since the electron and hole 
are indeed confined to the conduction and valence bands, respectively. Despite 
all the assumptions involved the energy spectrum that results from the 
one-band EMA model, when it is applied to a single electron, is surprisingly 
good, so this model finds its natural application in the modeling of electrons 
confined in  two- and three-dimensional electrostatically induced quantum dots, 
where the charging of the dot is achieved, for instance, lowering the 
electrostatic barriers that define the dot or connecting it to ``electron 
reservoirs'' through metallic leads, so neither an electron is promoted from 
the valence to the conduction band nor a exciton is formed.

A more sophisticated approach to deal with the band structure employs the 
$\mathbf{k}\cdot\mathbf{p}$
model \cite{bastard1990,voon2009} which, as the EMA,   depends on a number of 
effective 
parameters, but incorporates in a more natural fashion the crystal structure of 
the materials, the interaction between the energy bands, mechanical strain 
\cite{stier1999,winkelnkemper2006,jiang1997}, 
spin-orbit interaction and so on. Besides, the  results  calculated using the 
$\mathbf{k}\cdot\mathbf{p} $  are more accurate than those obtained using the 
EMA aproximation and closer to the experimental ones 
\cite{jaskolski1998,pryor1998}. It is important to note that the spectrum 
obtained 
using the method corresponds to the ``empty'' dot so, if the excitonic energies 
are to be calculated, the interaction between a particle occupying a level 
below the gap and other occupying a level above the gap must be introduced, 
again, assuming that it takes the form of a known potential 
\cite{li2000,kayanuma1988,takagahara1993}. The 
optical selection rules prevent that  all the excitons that can be formed 
could be excited in a direct fashion. The lowest lying exciton is called  
``dark exciton'' (DE) precisely for this reason, while the excitons that can be 
optically excited are conversely termed ``bright excitons'' 
(BE) \cite{rodina2016}. Depending on the nanostructure under study the DE 
recombination strongly modifies its photoluminescence properties at low 
temperatures. At any rate, the ground exciton state is studied on its own to 
characterize the physical traits of nanostructures or bulk materials in many 
different situations. The dark and bright excitons spectrum can be calculated 
effectively using the   $\mathbf{k}\cdot\mathbf{p}$ method by introducing an 
effective electron-hole interaction and constructing electron and hole wave 
functions with definite angular momentum quantum numbers \cite{rodina2016}.   

It is worth to mention that since excitons in semiconductor devices are less 
susceptible to perturbations produced by phonons than single electrons, once an 
exciton is formed its mean lifetime might be long enough 
to implement Quantum  Information tasks \cite{zrenner2002,krenner2005}. This 
fact has led to numerous attempts to identify systems where the coherent 
control, readout and initial preparations of excitonic states can be performed 
accurately and in a reliable way. 

There is not a reason to limit the number of layers to a core, a well and a 
barrier and it has been shown that spherical layered dots with two wells have 
some remarkable properties 
\cite{chang1998,ferron2012,ferron2013,holovatsky2017,holovatsky2016,karaca2016,
alharbi2016} , in 
particular the EMA model for a single electron 
confined in the dot predicts that  the wave 
function of few excited states can be located in the innermost well while the 
the wave function of all the other eigenstates is located in the outermost 
well \cite{ferron2013}. By located in a well it is understood that 
the 
probability that the electron is outside a given well is negligible. In 
Reference~\cite{ferron2013} it was shown that the number of eigenstates located 
in the innermost well can be tuned precisely changing the radius of the core, 
while in Reference~\cite{ferron2012} the localization phenomenon was studied 
only for the ground state.

In view of the results obtained using the EMA approach with respect to 
the localization properties of the electron wave function in spherical layered 
QD,  it is reasonable to wonder
about the behaviour of an exciton in this kind of structure when it is studied 
using the$\mathbf{k}\cdot\mathbf{p}$  model. 
Moreover, since the electron and the hole have different effective masses 
it is to be expected that their localization properties differ from what it is 
observed in the EMA approach for a single electron. The goal of the present 
study is to characterize the behaviour of an exciton bounded in a 
spherical layered 
quantum dot with two wells, using the $\mathbf{k}\cdot\mathbf{p}$  model to 
obtain the energy bands of the nanostructure. The manuscript is organized as 
follows, the modelling of spherical layered QDs and the 
$\mathbf{k}\cdot\mathbf{p}$  model are depicted in  Section~\ref{sec:modelo}, 
while the properties of the spectrum, eigenstates and excitons in structures 
with one or two wells are shown in Section~\ref{sec:resultados}. The spectrum 
and eigenfunctions are obtained using a high-precision Rayleigh-Ritz 
variational method. In Section~\ref{sec:conclusion} we summarize and 
discuss our results. Most technical details about the variational method, the 
variational basis set functions are referred to Appendices.

\section{Brief description of the layered QD 
structures and the $\mathbf{k}\cdot\mathbf{p}$ model}
\label{sec:modelo}

\begin{figure}[t]
\begin{center}
\includegraphics[width=0.8\textwidth]{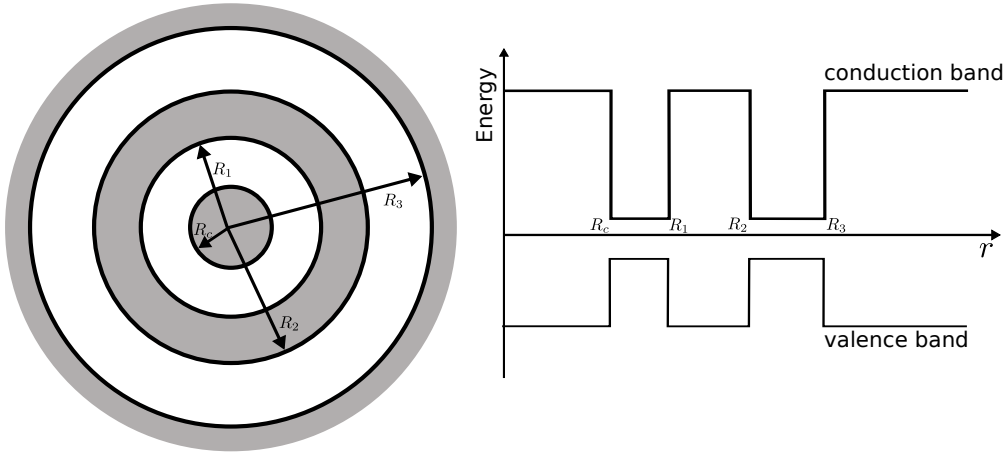}
\end{center}
\caption[Esquema de un punto cuántico doble y los respectivos perfiles de las 
bandas
de valencia y de conducción]{\label{fig_kp_fig5} The cartoon on the left 
depicts the cross-section of a  ZnS/CdSe/ZnS/CdSe/ZnS structure. A grey 
zone correspond to  a layer (or core) of  ZnS, while a white zone correspond to 
a layer of CdSe. The core radius, $R_c$ and the outermost radius of each layer, 
$R_1$, $R_2$ and $R_3$ are also shown.  The top of the valence band and the 
bottom of the conduction one are sketched as functions of the radial coordinate 
on the right side of the Figure.  }
\end{figure}

Besides the core, wells and barriers structure, some models of multilayered 
quantum dots include some cladding, usually a dielectric material with a band 
gap larger than the gap of the materials that form the quantum dot. This amounts 
for an homogeneous boundary condition for the wave functions. The model 
considered in this Section do not have a cladding, instead of that it is 
supposed that the outermost barrier extends indefinitely. In 
Figure~\ref{fig_kp_fig5} a cartoon is used to depict the layered structure 
associated to the quantum dot. The target-like graph depicts, using gray and 
light rings the barriers and wells present in the quantum dot, respectively, 
while the central disk corresponds to the core. The band structure profile is 
shown schematically as a function of the distance to the center of the QD. 
Along the work we use the known parameters of the semiconductors $ZnS$ and 
$CdSe$, this allows us to compare with results previously obtained using the 
EMA approach in Reference~\cite{ferron2013} for a ZnS/CdSe/ZnS/
CdSe/ZnS QD. Along this work the core radius is denoted by $R_c$, while the 
other radii will be denoted as $R_1$, $R_2$, $R_3$ and so on, see 
Figure~\ref{fig_kp_fig5}.

The $\mathbf{k}\cdot\mathbf{p}$ method has a huge number of variants that allow 
to include in the model  a great deal of information about the 
semiconductor materials used to construct a given QD. Because the materials  
included in the layered QD that we intend to study the 8-band model for a 
zincblende crystal structure is the natural choice, its Hamiltonian given by  
\cite{voon2009,kishore2012,kishore2014}

\begin{equation}\label{eq_kp_8bh}
H = \left(\begin{array}{cccccccc}
    A & 0 & V^{\star}  & 0 & \sqrt{3}V & -\sqrt{2}U & -U & \sqrt{2}V^{\star} \\
    0 & A & -\sqrt{2}U & -\sqrt{3}V^{\star} & 0 & -V & \sqrt{2}V & U \\
    V & -\sqrt{2}U & -P+Q & -S^{\star} & R & 0 & \sqrt{\frac{3}{2}}S & 
-\sqrt{2}Q \\
    0 & -\sqrt{3}V & -S & -P-Q & 0 & R & -\sqrt{2}R & \frac{1}{\sqrt{2}}S \\
    \sqrt{3}V^{\star} & 0 & R^{\star} & 0 & -P-Q & S^{\star} &
             \frac{1}{\sqrt{2}}S^{\star} & \sqrt{2}R^{\star} \\
    -\sqrt{2}U & -V^{\star} & 0 & R^{\star} & S & -P+Q & \sqrt{2}Q &
             \sqrt{\frac{3}{2}} S^{\star} \\
    -U & \sqrt{2}V^{\star} & \sqrt{\frac{3}{2}}S^{\star} & -\sqrt{2}R^{\star} &
             \frac{1}{\sqrt{2}}S & \sqrt{2}Q & -P-\Delta & 0 \\
    \sqrt{2}V & U & -\sqrt{2}Q & \frac{1}{\sqrt{2}}S^{\star} & \sqrt{2}R &
             \sqrt{\frac{3}{2}}S & 0 & -P-\Delta
    \end{array}\right) \,,
\end{equation}

\noindent where the entries in Equation (\ref{eq_kp_8bh}) are given by

\begin{eqnarray}
\label{eq_kp_eh1}
A &=& E_{c} + \frac{\hbar^2 k^2}{2m_0} \,,\\
\label{eq_kp_eh2}
P &=& -E_{v} + \gamma_1 \frac{\hbar^2 k^2}{2m_0} \,,\\
\label{eq_kp_eh3}
Q &=& \gamma_2 \frac{\hbar^2}{2m_0}\left(k_x^2 + k_y^2 - 2\,k_z^2\right) 
\,, \\
\label{eq_kp_eh4}
R &=& -\sqrt{3} \frac{\hbar^2}{2m_0}
      \left[\gamma_2\left(k_x^2 - k_y^2\right) - 2i\gamma_3 k_x k_y\right]\,,\\
\label{eq_kp_eh5}
S &=& \sqrt{3} \gamma_3 \frac{\hbar^2}{m_0}\,k_z\left(k_x - i k_y\right) \,,\\
\label{eq_kp_eh6}
U &=& \frac{i}{\sqrt{3}} P_0\, k_z\,,\\
\label{eq_kp_eh7}
V &=& \frac{i}{\sqrt{6}} P_0\, \left(k_x - i k_y\right)\,,
\end{eqnarray}

\noindent where $E_c$ is the energy at the bottom of the conduction band, $E_v$ 
the energy at the top of the valence band, so $E_g = E_c-E_v$ is the band gap, 
  $\gamma_1$, $\gamma_2$ y $\gamma_3$ are the Luttinger parameters, $\Delta$ is 
the intensity of the spin-orbit interaction, $P_0$ the interaction strength 
between both bands, and this quantity is related to the Kane's energy, $E_p$, 
by

\begin{equation}\label{eq_kp_Ep}
E_p = \frac{2\,m_0\,P_0^2}{\hbar^2}\,.
\end{equation}

For fixed values of $k_x,k_y$ and $k_z$ the Hamiltonian has eight eigenvalues, 
one for each sub-band. The largest pair belong to the conduction band, while 
the other six belong to the valence band and are the origin of the well known 
heavy-hole, light-hole, and split-off sub-bands. For a material without 
spin-orbit interaction each sub-band is doubly-degenerate, but in materials 
with a large spin-orbit interaction the degeneracy is broken.

As has been said above, the eight band Hamiltonian depends on the crystal 
structure, which is introduced via the Bloch functions, in our case the eight 
Bloch functions are given by

\begin{eqnarray}\label{eq:Bloch-function-uno}
u_{1/2}^c &=& S\uparrow, \\
u_{-1/2}^c &=& S\downarrow, \\
u_{3/2,3/2}^{v}  &=& \frac{1}{\sqrt{2}}(X+iY)\uparrow,\\
u_{3/2,1/2}^{v}  &=& \frac{i}{\sqrt{6}}[(X+iY)\downarrow-2Z\uparrow],\\
u_{3/2,-1/2}^{v} &=& \frac{1}{\sqrt{6}}[(X-iY)\uparrow+2Z\downarrow],\\
u_{3/2,-3/2}^{v} &=& \frac{i}{\sqrt{2}}(X-iY)\downarrow,\\
u_{1/2,1/2}^{v}  &=& \frac{1}{\sqrt{3}}[(X+iY)\downarrow + Z\uparrow],\\
u_{1/2,-1/2}^{v} &=& \frac{i}{\sqrt{3}}[-(X-iY)\uparrow + 
Z\downarrow],\label{eq:Bloch-function-ocho}
\end{eqnarray}

\noindent where the supraindexes $c$ and $v$ correspond to the 
conduction and valence band, respectively, and $S$ denotes a function that has 
spherical symmetry, while $X$, $Y$ and $Z$ are functions that combined have the 
same symmetry that $Y_{1m}$, {\em i.e.} the spherical harmonics with quantum 
number $\ell=1$., for instance $X+iY$ has the same symmetry as $Y_{11}$.  The 
parameters that enter in Equation~\ref{eq_kp_8bh} are matrix elements that must 
be calculated using  the functions in 
Equations~\ref{eq:Bloch-function-uno}--\ref{eq:Bloch-function-ocho} but, 
usually, are  
phenomenologically determined. Nevertheless,  since the Bloch basis gives the 
precise ordering of the matrix elements in Equation~\ref{eq_kp_8bh} we present 
them explicitly. 

To obtain the band structure of a given nano-structure $k_x,k_y$ and $k_z$ 
cannot be considered parameters, as is the case for the bulk. Instead, the 
transformation $k_j = -i\frac{\partial}{\partial x_j}$ with $j = x, y, 
z$ leads the matrix Hamiltonian in Eq.~\ref{eq_kp_8bh} to a set of eight 
coupled Schr\"odinger-like equations, whose eigenvalues give the band structure 
of the nanodevice. 

So, calling $\mathrm{H}$ the $8\times 8$ differential operator that is obtained 
from 
Eq.~\ref{eq_kp_8bh}, we look for the eigenvalues and eigenfunctions given by

\begin{equation}\label{eq_kp_sist_ec}
\mathrm{H}\, \mathbf{\Psi}(\mathbf{r}) = E\,\mathbf{\Psi}(\mathbf{r})\,,
\end{equation}

\noindent where $\mathbf{\Psi}(\mathbf{r})$ is a column vector 

\begin{equation}\label{eq_kp_fvec}
\mathbf{\Psi}(\mathbf{r}) = \left(\begin{array}{c}
                           \psi_1(\mathbf{r}) \\
                           \psi_2(\mathbf{r}) \\
                           \psi_3(\mathbf{r}) \\
                           \psi_4(\mathbf{r}) \\
                           \psi_5(\mathbf{r}) \\
                           \psi_6(\mathbf{r}) \\
                           \psi_7(\mathbf{r}) \\
                           \psi_8(\mathbf{r}) 
                           \end{array}
                         \right)\,.
\end{equation}

The parameters of the materials that form the QD's to be studied are given in 
Table~\ref{kp_tabla1} and Table~\ref{kp_tabla2}-

The band structure of a layered QD must be obtained solving the eigenvalue 
problem in Equation~\ref{eq_kp_sist_ec}, considering the parameters as spatial 
functions. For instance, if $r$, the radial coordinate, is such that $r<R_c$ 
then the parameters are taken as the ones of the $ZnS$, while  if $R_c<r<R_b$, 
then the parameters correspond to those of the $CdSe$ semiconductor, and so on. 
This is the case for the effective masses, the energy gap, the Kane's energy, 
the Luttinger parameters, etc.  The discontinuity introduced in the material  
parameters, particularly for the masses, suggests that some care must be 
exercised to ensure that the problem is Hermitian \cite{haus1993,ferron2013} 
and that this care must be 
extended to the matching conditions for the functions $\psi_i$ at the 
interfaces between two different semiconductors . 

In the Appendices~\ref{ap:b-splines}, \ref{ap:basis-b-splines} and \ref{ap:kp} 
we briefly present the necessary details to implement a variational calculation 
of the eigenvalues and eigenfunctions, Equation~\ref{eq_kp_sist_ec}, using a 
B-splines basis set \cite{Boor1978,Bachau2001}. Some extra steps can be found 
in 
\cite{ferron2013,Garagiola2018} and References therein. 

There are several forms to deal with the interaction between the electron-hole 
pair, but the two that atract more attention deal with it assuming a 
Coulomb-like potential 
\cite{schooss1994,chang1998,billaud2009,li2000,ferreyra1999} or consider that 
the interaction is the solution of the Poisson equation within some 
approximation. For instance, within the Hartree approximation 
\cite{yao2010,stier1999} it 
is assumed that the 
wave function of the pair is given by $\Psi(\mathbf{r}_e,\mathbf{r}_h) =
\psi_e(\mathbf{r}_e)\psi_h(\mathbf{r}_h)$ where $\psi_e(\mathbf{r}_e)$ y
$\psi_h(\mathbf{r}_h)$ are the wave functions of the electron and hole, 
respectively. Both one-particle wave functions are self-consistently calculated 
solving the eigenvalue problems

\begin{eqnarray}
\label{eq_kp_hartree1}
\left[H + V_e\right]\psi_h &=& \tilde{E}_h\,\psi_h\,,\\
\label{eq_kp_hartree2}
\left[H + V_h\right]\psi_e &=& \tilde{E}_e\,\psi_e\,,
\end{eqnarray}

\noindent where $\tilde{E}_e$ and $\tilde{E}_h$ are eigenvalues that belong to 
the conduction and valence bands, respectively. The electrostatic potentials 
$V_e$ and $V_h$ are the solutions 
of a pair of Poisson equations

\begin{eqnarray}
\label{eq_kp_poisson1}
-e|\psi_e|^2 &=& \epsilon_0\nabla\left(\epsilon_s(\mathbf{r})\nabla 
V_e\right)\,,\\
\label{eq_kp_poisson2}
 e|\psi_h|^2 &=& \epsilon_0\nabla\left(\epsilon_s(\mathbf{r})\nabla 
V_h\right)\,.
\end{eqnarray}

\noindent The potentials $V_e$ and $V_h$ are continuous functions, but at the 
interfaces between two material the normal derivative is discontinuous because 
of the different dielectric constants of the materials at both sides of a
interface. The problem of the polarization charges induced at the interface 
surfaces has been treated extensively 
\cite{takagahara1993,ougut1997,banyai1992,ferreyra1998,winkelnkemper2006}. The 
calculation method that is used to obtain the results reported in the next 
Section, a variational approach using $B$-splines functions as basis set, 
allows us to introduce the matching conditions for the functions 
$\psi_i$ {\em and} the potentials explicitly. So, in our results the effects of 
the polarizations charges is included in the solutions of the Poisson 
Equations~\ref{eq_kp_poisson1} and \ref{eq_kp_poisson2}.

In Equations \ref{eq_kp_hartree1} and \ref{eq_kp_hartree2} $H$ is the eight 
band Hamiltonian, and $\epsilon_s(\mathbf{r})$ is the dielectric constant as a 
function of the position along the radial coordiate of the QD. 

The system of equations formed by 
Equations~\ref{eq_kp_hartree1}, \ref{eq_kp_hartree2}, \ref{eq_kp_poisson1} and 
\ref{eq_kp_poisson2} is  solved iteratively, using as a starting point the 
solutions found for the ``empty'' QD, Equation~\ref{eq_kp_sist_ec}, to 
calculate the potentials $V_e$ and $V_h$ through 
Equations~\ref{eq_kp_poisson1}, \ref{eq_kp_poisson2} and then these potentials 
are introduced in Equations~\ref{eq_kp_hartree1},\ref{eq_kp_hartree2}. New wave 
functions are calculated, which are then used to obtain new potentials and so 
on, until the energies obtained are stabilized. Again, the details of the 
procedure are deferred to the Appendices. In the following we denote as 
 $\psi_{e}^0$ and $\psi_{h}^0$ the eigenfunctions of the empty QD that 
correspond to eigenvalues above and below the energy gap and that will be 
used as the starting point of the iterative solution of 
Equations~\ref{eq_kp_poisson1},\ref{eq_kp_poisson2}. {\em A priori} any pair of 
electron $E_e,\psi_e^0$, and hole $E_h,\psi_h^0$ energies and 
eigenfunctions could be used to calculate exciton binding energies but, most 
commonly, the calculation is restricted to the lowest and highest 
eigenvalues on the conduction and valence bands, respectively, and their 
respective eigenfunctions or, in other words, we focus in the 
dark exciton energy.

Finally, once stabilized values of the energies $\tilde{E}_h$ and $\tilde{E}_e$ 
are obtained, the binding energy can be calculated as

\begin{equation}\label{eq_kp_binding_E}
E_b = \left(E_e - E_h\right) - \left(\tilde{E}_e - \tilde{E}_h\right) =
\left(E_e - \tilde{E}_e\right) + \left(\tilde{E}_h -E_h\right)\,,
\end{equation}

\noindent where $E_e$ and $E_h$ are energy values taken from the spectrum in 
Equation~\ref{eq_kp_sist_ec} that where used, together with their respective 
eigenfunctions, as the first step in the iteration procedure described above. 
Most commonly, $E_e$ is taken as the lowest energy eigenvalue  in the 
conduction band, while $ E_h$ is the highest energy eigenvalue in the valence 
band. Proceeding along these lines, $(E_c-E_h)-E_b$ is the lowest energy that a 
photon should have to promote an electron from the valence to the conduction 
band.

\section{Results}\label{sec:resultados}

\subsection{A QDQW structure made of CdS/HgS/CdS }\label{sec:one-well}

To check the correct implementation of the different algorithms a QDQW 
structure with a $CdS$ core and barrier, and a HgS well was studied thoroughly. 
The election of the materials allows us to compare with results previously 
obtained using the EMA approach. All the parameters used to determine the 
$\mathbf{k}\cdot\mathbf{p}$ Hamiltonian can be found in Table~\ref{kp_tabla1}.

\setlength{\tabcolsep}{0.5em}
{\renewcommand{\arraystretch}{1.2}
\begin{table}[h]
\caption{\label{kp_tabla1} Parámetros de los materiales $CdS$ y $HgS$ utilizados
en el punto cuántico.}
\begin{center}
\begin{tabular}{c c c c c c c c}
\hline \hline
       & $E_g$ & $E_p$ & $\Delta$ & $\epsilon$ & $\gamma_1$ & $\gamma_2$ & 
$\gamma_3$ \\
\hline
$CdS$  & 2.42 $eV$ & 19.6 $eV$ & 0.08 $eV$ & 5.5   & 0.814 & 0.307 & 0.307 \\
$HgS$  & 0.42 $eV$ & 21.0 $eV$ & 0.08 $eV$ & 11.36 & 12.2  & 4.2   & 4.2 \\
\hline \hline
\end{tabular}\\
\vspace{0.3cm}
CBO = 1.35 $eV$ \\
VBO = 0.65 $eV$
\end{center}
\end{table}
}

\noindent In the Table above CBO stands for the {\em conduction band offset}, 
{\em i.e.} 
the 
energy difference between the bottom of the conduction bands of both materials, 
while VBO stands for the  {\em valence band offset}, which is the difference 
between the top of the valence bands for both materials.

\begin{figure}[t]
\begin{center}
\includegraphics[width=0.85\textwidth]{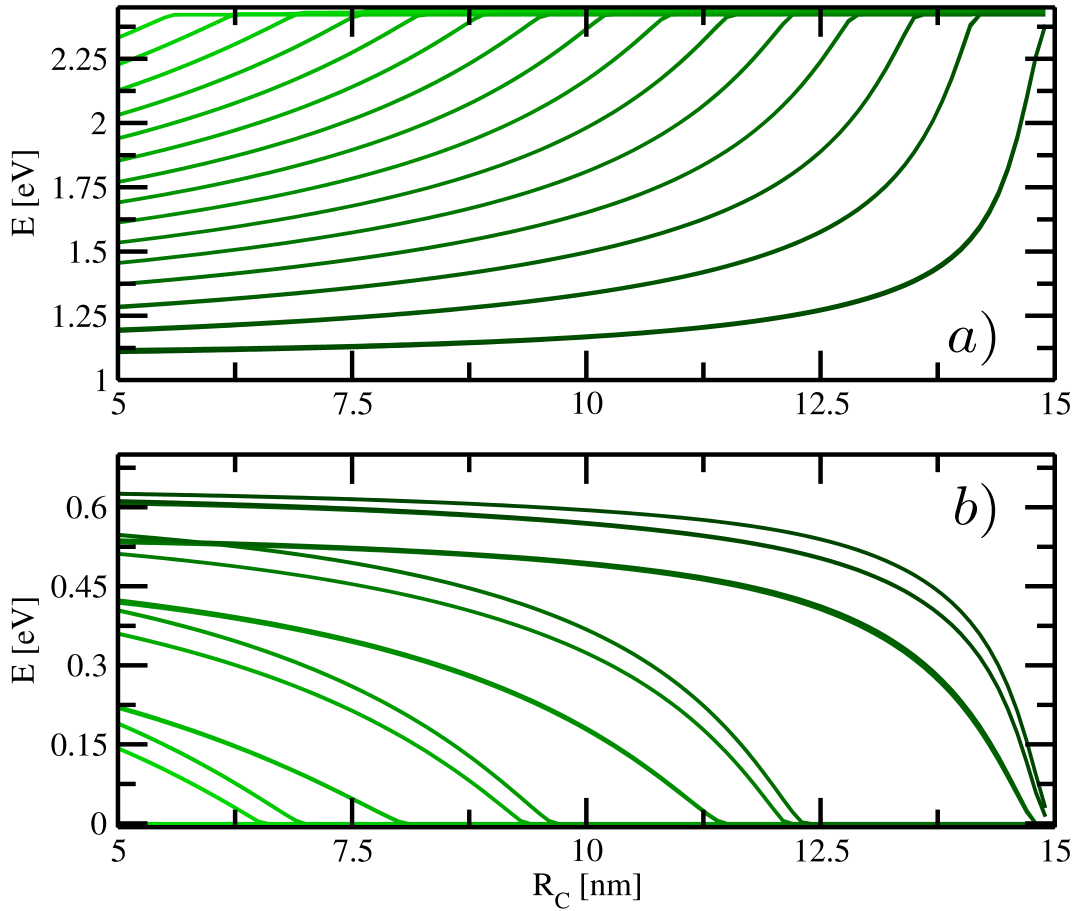}
\end{center}
\caption[Espectro de energía de un punto cuánticos esférico en el modelo
$\mathbf{k}\cdot\mathbf{p}$]{\label{fig_kp_fig2} The variational eigenvalues 
calculated for the spherical QDQW structure CdS/HgS/CdS, for both the EMA 
(green curves) and $\mathbf{k}\cdot\mathbf{p}$ (black curves) approaches. a) 
Energy levels on the conduction band. b)  Energy levels on the valence band.}
\end{figure}

\begin{figure}[ht]
\begin{center}
\includegraphics[width=0.85\textwidth]{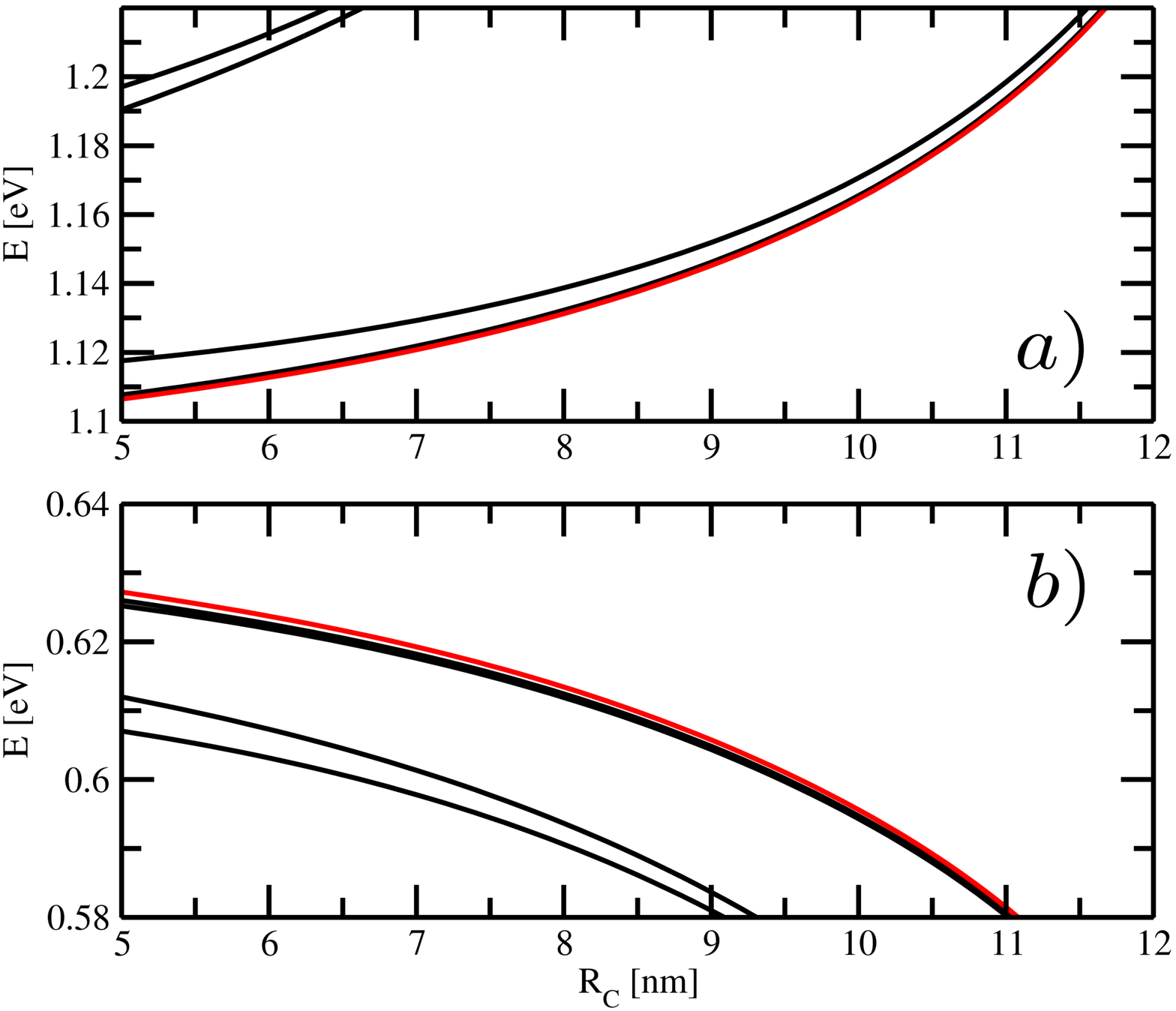}
\end{center}
\caption[Espectro de energía de un punto cuántico esférico cuando se consideran 
la
interacción entre un electrón y un hueco.]{\label{fig_kp_fig3}$a)$Eigenvalues 
on the conduction band. $b)$ eigenvalues on the valence band. Both panels show 
a few eigenvalues that lie near the energy gap corresponding to the empty QD 
(black curves). The eigenvalues obtained implementing the Hartree approximation 
for the lowest eigenvalue of the conduction band and the higher one of 
the valence band (and corresponding eigenstates) are shown as red curves in 
both panels. It is clear that the energy gap predicted by the Hartree 
approximation is smaller than the gap of the empty QD.  }
\end{figure}

\begin{figure}[ht]
\begin{center}
\includegraphics[width=0.6\textwidth]{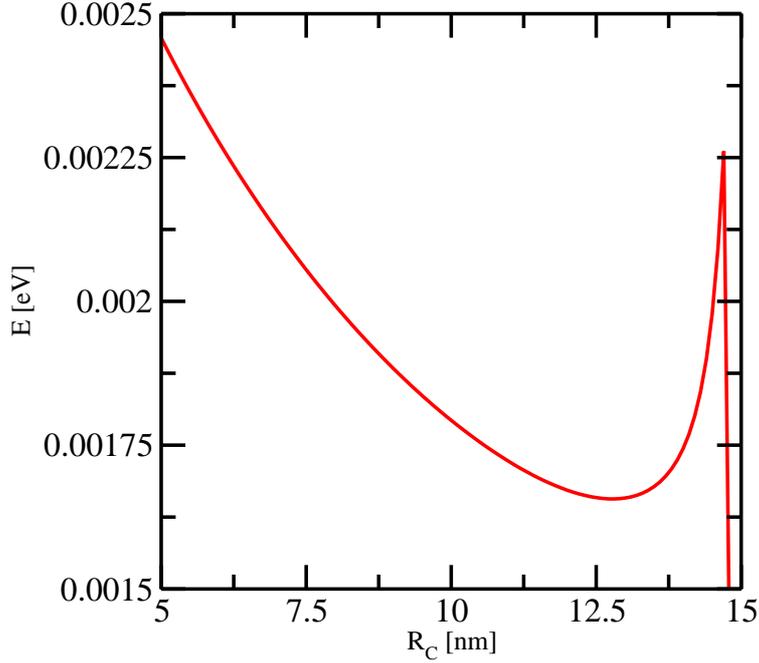}
\end{center}
\caption[Energía de ligadura de un excitón en un punto cuántico con simetría
esférica]{\label{fig_kp_fig4}The binding energy calculated for an exciton 
trapped in a QDQW structure using the Hartree approximation together with 
the $\mathbf{k}\cdot\mathbf{p}$ method. The electron and hole 
states considered are those corresponding to the lowest and highest eigenvalue 
on the conduction and valence bands, respectively.}
\end{figure}

The results for the energy spectrum, the Hartree energy levels and the binding 
energy are summarized in Figures~\ref{fig_kp_fig2}, \ref{fig_kp_fig3} and 
\ref{fig_kp_fig4} respectively. The structure studied has a core with radius 
$R_c$ and the well is characterized by two radii, the innermost one $R_c$ and 
the outermost one, $R_1=15$ nm. 

Figure~\ref{fig_kp_fig2}a)  shows the lowest lying variational eigenvalues 
calculated using the  $\mathbf{k}\cdot\mathbf{p}$ and EMA models {\em vs} the 
core radius. The number of bound states decreases when the core radius 
increases. The sets of eigenvalues calculated using the EMA and the
$\mathbf{k}\cdot\mathbf{p}$ method  are remarkably close which, to some 
extent, 
is to be expected since the interaction between bands and other effects that 
the $\mathbf{k}\cdot\mathbf{p}$ method takes into account (and the EMA 
method not) are relatively weak for the materials and QD dimensions considered. 
The 
valence band eigenvalues, shown in  Figure~\ref{fig_kp_fig2}b), also for both 
the $\mathbf{k}\cdot\mathbf{p}$ and EMA models, behave similarly to those in 
the conduction band. An appreciable difference can be observed for small 
core 
radius where the spectrum obtained with the $\mathbf{k}\cdot\mathbf{p}$ model 
shows some avoided crossings, signaling that for those $R_c$ values the 
interaction between the different bands is large enough. On the other hand, the 
eigenvalues obtained from the EMA are smooth and can not show any avoided 
crossing see, for instance Reference~\cite{ferron2013} and References therein.  

An important difference between the EMA approach and the 
$\mathbf{k}\cdot\mathbf{p}$ method arises because the Hamiltonian of the 
second, Equations~\ref{eq_kp_8bh}, has terms that do not commute with the 
orbital angular momentum operator, $\mathbf{L}^2$, see 
Equations~\ref{eq_kp_eh4}, \ref{eq_kp_eh5},\ref{eq_kp_eh6} and  
\ref{eq_kp_eh7}. This has at least two consequences, the first has to do with 
the labeling of the eigenfunctions and eigenvalues in both methods. In the EMA, 
the inclusion of orbital angular momentum has the only effect that the radial 
potential changes and that the quantum numbers of the eigenfunctions and 
eigenvalues  are the radial and the angular momentum ones, this is not true for 
the eigenfunctions and eigenvalues obtained using the 
$\mathbf{k}\cdot\mathbf{p}$ Hamiltonian. Second, the basis set necessary to 
implement a variational approximation for the EMA approach has functions that 
only depend on the radial coordinate, while the basis set of to implement the 
variational calculation in the $\mathbf{k}\cdot\mathbf{p}$ approach requires 
functions that depend on the radial and angular coordinates, see 
Appendix~\ref{ap:kp}. The second argument explains why it is easier to obtain a 
larger number of bounded eigenfunctions with the EMA approach than with the 
$\mathbf{k}\cdot\mathbf{p}$ approach. The energy of the eigenvalues in the EMA 
approach strongly depends  on the radial quantum number but more weakly with 
the angular momentum quantum number (see Reference~\cite{ferron2013}.

Figure~\ref{fig_kp_fig3} shows the Hartree eigenvalues $\tilde{E}_h$ and 
$\tilde{E}_e$, Equations~\ref{eq_kp_hartree1} and \ref{eq_kp_hartree2}, 
corresponding to the highest and lowest eigenvalues of the valence and 
conduction band, respectively, while Figure~\ref{fig_kp_fig4} shows the binding 
energy {\em vs} the core radius, $R_c$.

The iterative procedure to solve simultaneously the Poisson equations 
and the $\mathbf{k}\cdot\mathbf{p}$ Hamiltonian converges rapidly, for all the 
values shown in Figure~\ref{fig_kp_fig4}a) and b) the procedure converged after 
less than ten iterations. As it is to be expected, the  interaction between the 
hole and the electron leads to a positive correction for the eigenvalue located 
in the valence band, and a negative one for the conduction band eigenvalue.

The behaviour of the binding energy, shown in Figure~\ref{fig_kp_fig4} merits 
a detailed analysis. At least since the work of Bryant \cite{bryant1995}, it is 
well known that the energy of an exciton decreases with the width of the 
potential well. Using perturbation theory, it is rather direct to show that the 
exciton binding energy decays as $1/L^{\alpha}$, where $L$ is the width of the 
well in the QDQW structure, and $\alpha$ is some exponent. If the interaction 
between the hole and electron pair is given by a simple Coulomb term, $\alpha$ 
is, with a good accuracy, a natural number. Otherwise, if the interaction 
between the electron and the hole is not included only in terms of a 
Coulomb-like term, and the effects of polarization charges are taken into 
account the simple decaying behaviour can not be observed. In a QDQW structure 
there are several parameters that can be changed, we choose to keep the 
external well radius fixed and change the core radius, $R_c$, so when the core 
radius decreases the actual width of the well is being increased. This explains 
why the binding energy grows when $R_c$ goes from 12.5 nm to 15 nanometers. For 
From $R_c\simeq 12.5$ to $R_c\rightarrow 0$ the binding energy grows when the 
width of the potential well increases. This behaviour is owed to the non-linear 
effects of the polarization charges and the change in the boundary condition at 
the center of the QD core.

\subsection{A QD structure made of ZnS/CdSe/ZnS/CdSe/ZnS}\label{sec:two-wells}

\begin{figure}[ht]
\begin{center}
\includegraphics[width=0.85\textwidth]{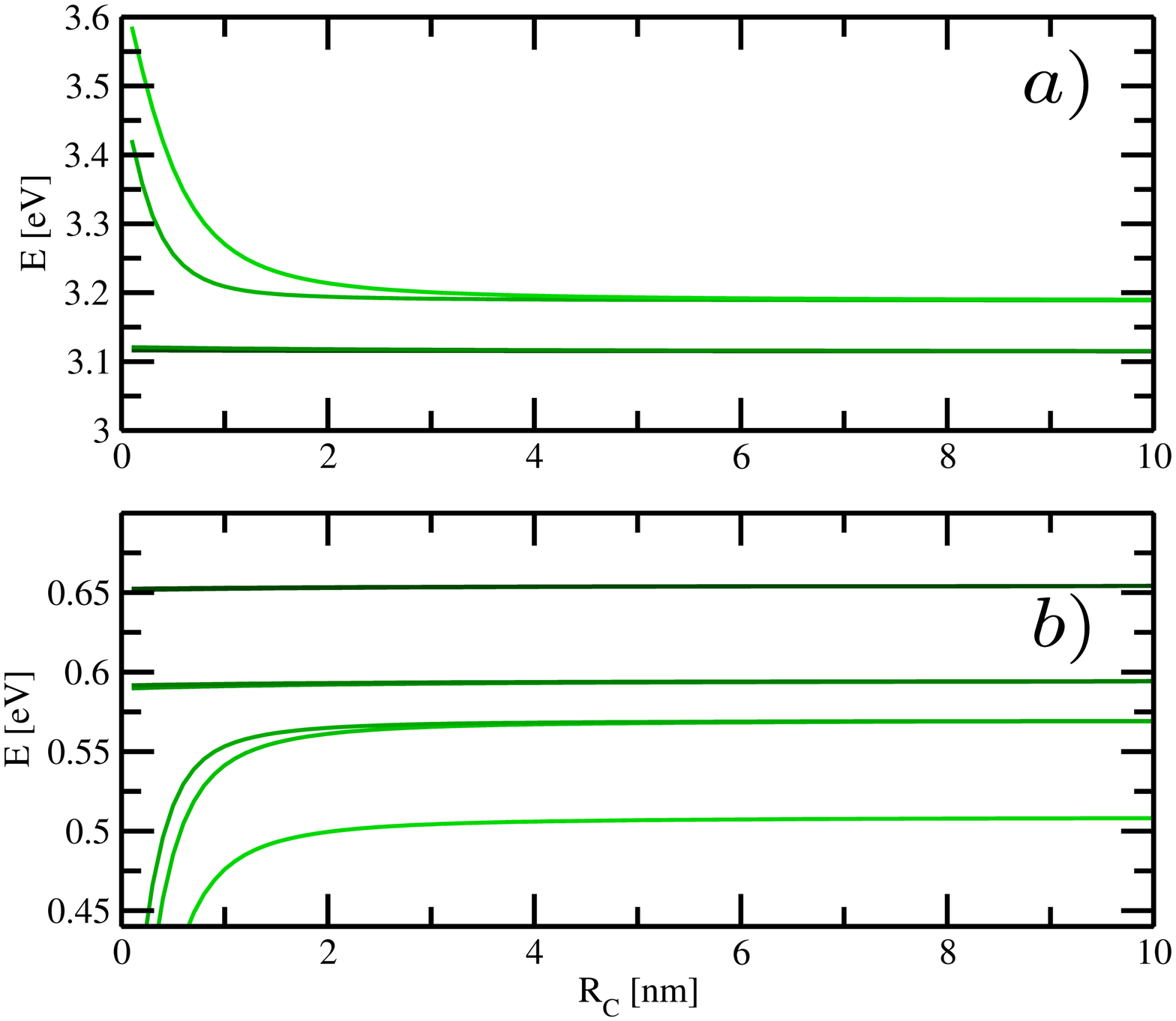}
\end{center}
\caption[Espectro de energía de un punto cuántico 
doble]{\label{fig_kp_fig6} The variational eigenvalues 
calculated for the spherical structure ZnS/CdSe/ZnS/CdSe/ZnS, for both the 
EMA 
(green curves) and $\mathbf{k}\cdot\mathbf{p}$ (black curves) approaches. a) 
Energy levels on the conduction band. b)  Energy levels on the valence band.}
\end{figure}

In the following a QD with a core made of $ZnS$ and two wells made of CdSe is 
considered. The corresponding parameters for both semiconductors can be foun in 
Table~\ref{kp_tabla2}- The materials and dimensions of the QD are chosen so to 
allow  a straightforward comparison with previous results \cite{ferron2013}, so 
in this case the width of the potential barrier that lies between the two 
potential wells and the widths of the two potential wells are kept fixed, while 
the core radius is used to tune the properties of the QD. In particular we 
choose $R_1 =R_c + 0.8$ nm; $R_2 = R_1 + 3.5$ nm, and $R_3 = R_2 + 1$ nm.

\setlength{\tabcolsep}{0.5em}
{\renewcommand{\arraystretch}{1.2}
\begin{table}[h]
\caption{\label{kp_tabla2} Parámetros de los materiales $ZnS$ y $CdSe$ 
utilizados
en el punto cuántico doble.}
\begin{center}
\begin{tabular}{c c c c c c c c}
\hline \hline
& $E_g$ & $E_p$ & $\Delta$ & $\epsilon$ & $\gamma_1$ & $\gamma_2$ & $\gamma_3$ 
\\
\hline
$ZnS$  & 3.68 $eV$ & 20.4 $eV$ & 0.074 $eV$ & 5.7  & 2.12 & 0.51 & 1.56 \\
$CdSe$ & 1.75 $eV$ & 17.4 $eV$ & 0.24 $eV$  & 6.3 & 4.4 & 1.6 &  2.68 \\
\hline \hline
\end{tabular}\\
\vspace{0.3cm}
CBO = 0.9 $eV$ \\
VBO = 1.03 $eV$
\end{center}
\end{table}
}

Figure~\ref{fig_kp_fig6}a and b) show the lowest lying variational eigenvalues, 
obtained with the $\mathbf{k}\cdot\mathbf{p}$, on the conduction band and the 
highest ones on the valence band, as functions of the core 
radius. The Figure also show the corresponding values obtained using the EMA. As 
it is in the case of a QDQW, both sets of values are quite similar. The 
proximity of the eigenvalues could lead to the conclusion that it is not 
necessary to  go through the complications of the $\mathbf{k}\cdot\mathbf{p}$. 
But, as the first example analyzed in this Section shows, for small values of 
$R_c$ the  $\mathbf{k}\cdot\mathbf{p}$ model predicts some features that can 
not be found in the results obtained from the EMA approach.

\begin{figure}[ht]
\begin{center}
\includegraphics[width=0.9\textwidth]{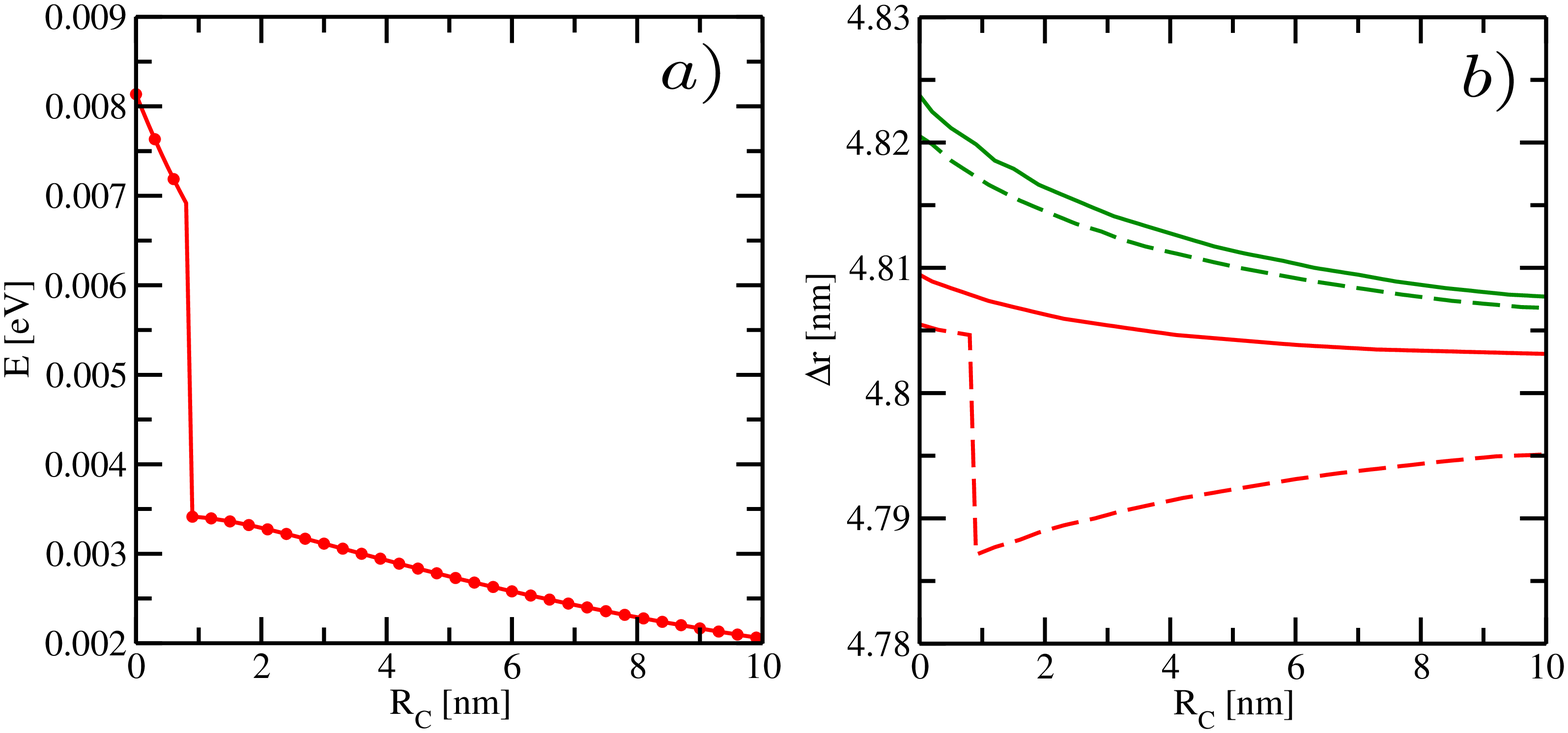}
\end{center}
\caption[Energía de ligadura del estado fundamental del excitón en punto 
cuántico
doble y posición del máximo de las densidades de probabilidad del electrón y
del hueco]{\label{fig_kp_fig7}$a)$ The binding energy calculated for an exciton 
trapped in a  structure with a double well, using the Hartree approximation 
together with 
the $\mathbf{k}\cdot\mathbf{p}$ method. The electron and hole 
states considered are those corresponding to the lowest and highest eigenvalue 
on the conduction and valence bands, respectively. $b)$ The value at which the 
probability distribution attains its maximum {\em vs} the core radius. The red 
dashed and continuous curves correspond to the electron probability 
distributions calculated for the empty QD and the Hartree approximation, 
respectively. The green curves correspond to the hole probability distributions 
and follow the same convention than the electronic ones. }
\end{figure}

Even though the spectrum shown in Figure~\ref{fig_kp_fig6} is, to some extent, 
quite featureless, the binding energy of the exciton that can be formed from 
the lowest and highest eigenvalue from the conduction and valence band, 
respectively, show a rather steep change around $R_c\simeq 1$ nm, see 
Figure~\ref{fig_kp_fig7}a). The effect can be qualitatively analyzed following 
the radii values where the probability distributions of the electron and hole 
attain their maximum, this quantity is shown in Figure~\ref{fig_kp_fig7}b).

The device under study is such that the outermost potential well is located 
between the radial coordinates $R_c+4.3$ nm and $R_c+5.3$ nm. So the data 
shown in Figure~\ref{fig_kp_fig7} shows that for $R_c \gtrsim 1$nm the maximum 
of the electronic probability distribution is near the inner interface of the 
second potential well ({\em i.e} near $R_c+4.3$ nm), while for $R_c \lesssim 
1$nm the maximum is located nearer to the interface at $R_c+5.3$ nm. The 
maximum of the hole probability distribution is always located near the 
outermost interface. This simple picture explains why the binding energy 
changes, and grows, for $R_c\lesssim 1$ nm. Of course, the change in the 
position of the electronic maximum must balance the effect of the polarization 
charges induced by the hole and electron at the interfaces, the boundary 
condition at the origin of the radial coordinate and the attractive force 
between the two particles.

\section{Discussion and Conclusions}\label{sec:conclusion}

The study of spherical layered  quantum dots using the 
$\mathbf{k}\cdot\mathbf{p}$ model remains mired by several technical 
difficulties: it is a coupled system of differential equations with many 
position-dependent parameters, and a number of matching conditions at the 
interfaces between the different materials. Moreover, if the  exciton spectrum 
is to be calculated then a pair of coupled Poisson equations, which have their 
own 
set of matching conditions, must be added and solved simultaneously together 
with those of the $\mathbf{k}\cdot\mathbf{p}$ model. The $B$-spline basis set 
allows to calculate a variational approximation to the eigenvalues and 
eigenvectors, together with the electrostatic potentials and  take into 
account all the matching and boundary conditions in an efficient way. The 
algebraic problem is sparce because of the ``orthogonality'' relationship 
between the $B$-splines, but its size grows quite fast. For instance, if the 
number of spline functions is $N$, the Hamiltonian  to be diagonalized is 
a $8 \ell N \times 8\ell N$ matrix, where $\ell$ is the number of angular 
functions used to construct the variational basis, see Appendix~\ref{ap:kp}. In 
the two QD studied $N\simeq 120$, so the Hamiltonian matrices considered were, 
approximately, $1900\times 1900$ matrices. The number of basis set functions 
used  in the variational calculations is 
large enough to ensure that the binding energy values shown in 
Figure~\ref{fig_kp_fig7} have relative errors of less than 5\%.  

The steep change in the binding energy, observable in Figure~\ref{fig_kp_fig7}, 
is owed to the Coulomb interaction and the polarization charges induced at the 
interfaces between the different materials from which the quantum dot is made. 
Since the change is related to changes in the localization of the 
electronic probability 
distribution it is possible that the same abrupt change can be produced 
applying external fields to the QD while keeping the core radius fixed. Anyway, 
if a external field is applied the symmetry of the problem is different and 
another numerical approach is required. In the same sense, to enhance the jump 
on the binding energy the quantum dot should be constructed with semiconductor 
materials with a large difference between their dielectric constants but we 
preferred to study  combinations of materials for the potential wells and 
barriers that are actually constructed.

Finally, to obtain the bright and dark excitons spectrum an 
approach similar to the used in \cite{rodina2016}, {\em i.e.} using an 
effective electron-hole interaction that allows to control the angular momentum 
quantum numbers of the electron and hole wave functions, is more direct than 
the Hartree approach used in this paper. Nevertheless the use of the $B$-spline 
basis functions is quite flexible and allows the treatment of problems where 
the exact radial wave is not available (or too convoluted) and the matching 
conditions of the wave functions at the material interfaces must be taken into 
account.  We are currently working to implement an approach that computes 
the exciton spectrum using the effective electron-hole interaction and the 
$B$-spline variational calculation for layered quantum dots.

\section*{Acknowledgments}
We acknowledge SECYT-UNC and CONICET for partial financial support of this 
project.

\appendix

\section{The  \texorpdfstring{$B$}{B}-splines functions}\label{ap:b-splines}

The $B$-splines functions are a generalization of the simple polynomial 
functions and were designed specifically as a basis to expand other functions. 
Thorough presentations of these functions and  their properties can be found in 
\cite{Bachau2001} and in \cite{Boor1978}

Let us introduce some definitions

\begin{itemize}

\item An order $k$ polynomial is given by
\[
 p(x) = a_0 + a_1 x + \ldots + a_{k-1} x^{k-1} ,
\]
{\em i.e.} it is a $k-1$ degree polynomial.

\item A sequence of points, in ascending order and not necessarily different, 
in the interval $[a,b]$ is called $knots$

\begin{equation*}
\{t_i\}_{i=1,\ldots,n_k}\,,\quad\quad a=t_1\leq t_2\leq \ldots \leq t_{n_k}=b\,.
\end{equation*}

\noindent A given sequence can be grouped accordingly with which elements are 
different as follows

\begin{equation*}
\begin{array}{l}
 t_1 = t_2 = \ldots = t_{\mu_1} = \zeta_1\,,\\
 t_{\mu_1+1} = t_2 = \ldots = t_{\mu_1+\mu_2} = \zeta_2\,,\\
 \vdots \\
 t_{n+1} = \ldots = t_{n+\mu_{l+1}} = \zeta_{l+1}\,,\,\,\,n=\mu_1 + \ldots + 
\mu_l\,.
\end{array}
\end{equation*}

\noindent The points $\zeta_j$ satisfy that
$\zeta_j < \zeta_{j+1}$ and split the interval $[a,b]$ in $l$
subintervals $I_j = [\zeta_j,\zeta_{j+1}]$, and this new sequence is termed 
``of breakpoints''.

\end{itemize}

All in all, each $B$-spline is a piecewise function, defined over adjacent 
subintervals. In each subinterval, the $B$-spline is given by a polynomial of a 
certain order $k$, and satisfies a given matching condition at the point where 
two subintervals meet. So, a set of $B$-splines are determined by the order of 
the polynomials and the particular sequence of $knots$ used in the interval 
$[a,b]$. The multiplicity of each $knots$ prescribes the differentiability of 
the function in that point. A multiplicity equal to the unity results in 
functions that are $C^{k-2}$. 

The more common choice for the sequence of $knots$ is obtained taken $knots 
\in (a,b)$ with multiplicity equal to the unity and a $knot$ in $a$ and $b$ 
with multiplicity equal to $k$. With this choice, the number of $B$-splines 
functions is given by

\begin{equation*}
n_b = l + k -1\,.
\end{equation*}

\noindent In the most general case, such that there are $n_k$ $knots$ some of 
them repeated, the number of $B$-spline functions is given by

\begin{equation*}
n_b = n_k - k\,.
\end{equation*}

--------
\section{\texorpdfstring{$B$}{B}-splines as a basis 
set to expand functions}\label{ap:basis-b-splines}

Once the order of the polynomials, $k$, and the sequence of the $knots$, 
$\{t_i\}_{i=1,\ldots,n_k} \in [a,b]$ are chosen, the set of $B$-splines is 
wholly determined. In what follows,  note  that we denote these 
functions by $B_j(x)$ or by $B_j^k$ when there is an equation that involves 
$B$-spline functions of the same order or several different ones, since in the 
last case the order must be addressed explicitly . To construct the basis set 
it is useful to state explicitly the following properties

\begin{itemize}
\item In each interval $(t_i,t_{i+1})$, there are only $k$ 
different from zero $B$-splines functions

\begin{equation}
  B_j(x) \neq 0\,\quad\mbox{for}\quad j=i-k+1,\ldots,i\,.
\end{equation}

\noindent as a consequence

\begin{equation}\label{eq:b-splines-orthogonality}
  B_i(x)\, B_j(x) = 0 \quad\mbox{for}\quad |i-j|\geq k\,.
\end{equation}

\item For all $x\in [a,b]$
\begin{equation}
 \sum_{i} B_i(x) = 1\,.
\end{equation}

 \item The $B$-splines of order $k$ can be constructed using a recurrence 
relationship in terms of $B$-spline functions of smaller orders

 \begin{equation}\label{eq:recurrencia_splines}
  B_i^k(x) = \frac{x-t_i}{t_{i+k-1}-t_i} B_i^{k-1}(x) +
             \frac{t_{i+k}-x}{t_{i+k}-t_{i+1}} B_{i-1}^{k-1}(x)\,,
 \end{equation}

\noindent which, together with the definition of $B$-splines of order 
$k=1$

\begin{equation}\label{eq:splinek1}
   B_i^1(x) = \left\{\begin{array}{c c}
                     1 & \mbox{if}\quad t_i\leq x\leq t_{i+1} \\
                     0 & \mbox{otherwise}
                     \end{array}\right.\,,
\end{equation}

\noindent provide the algorithm used to evaluate the $B$-splines at a given 
 $x$ value. In the book of de Boor \cite{Boor1978} there is a number of Frotran 
routines that implement the algorithm implied by Equations 
\ref{eq:recurrencia_splines} and \ref{eq:splinek1}.

\end{itemize}

\section{The variational basis set used to calculate the approximate spectrum 
of a multilayered QD}
\label{ap:kp}

The spectrum of a given QD is obtained by solving 

\begin{equation}\label{eq_kp_se}
\mathrm{H} \mathbf{\Psi}(\mathbf{r}) = E\,\mathbf{\Psi}(\mathbf{r})\,,
\end{equation}

\noindent where 

\begin{equation}\label{eq_kp_wf}
\Psi(\mathbf{r}) = \left(\begin{array}{c}
\psi_1(\mathbf{r}) \\
\psi_2(\mathbf{r}) \\
\psi_3(\mathbf{r}) \\
\psi_4(\mathbf{r}) \\
\psi_5(\mathbf{r}) \\
\psi_6(\mathbf{r}) \\
\psi_7(\mathbf{r}) \\
\psi_8(\mathbf{r}) 
\end{array}\right)\,,
\end{equation}

\noindent where $\mathrm{H} = H(k_i \rightarrow -i\frac{\partial}{\partial 
x_i})$, and $H$ is the eight band $\mathbf{k}\cdot\mathbf{p}$ Hamiltonian. 
Besides, we denote, for latter use, the matrix entries of the $8\times 8$ 
differential operator in Equation~\ref{eq_kp_se} as $\mathrm{H}^{i,j}$ 
($i,j=1,2,\ldots, 8$). 

A very accurate approximation for a number of 
eigenvalues can be obtained using the Rayleigh-Ritz variational model. 
Because the symmetry of the problem, and 
knowing that the the atomic levels that originate the conduction and valence 
bands have orbital angular momentum quantum number $\ell=0$ and $\ell=1$, 
respectively, we choose a basis set with functions

\begin{equation}\label{eq_basekp}
\phi_{n,l}(\mathbf{r}) = B_{n}(r)\,P_l(\cos(\theta))\,,
\end{equation}

\noindent where the $B$-spline functions depend on the spherical radial 
coordinate, $P_l(\cos \theta)$ are the Legendre polynomials. The variational 
functions $\psi_i^v(\mathbf{r}$ that approximate the functions in 
Equation~\ref{eq_kp_wf} are given by

\begin{equation}\label{eq_componentes}
\psi_i^v(\mathbf{r}) = \sum_{n,l} c_{n,l}^i\,\phi_{n,l}(\mathbf{r}),\quad 
i=1,\cdots,8\,.
\end{equation}

The variational spectrum is calculated solving the {\em matricial} eigenvalue 
problem

\begin{equation}\label{eq:generalized-variational-eigrnvalue-problem}
[\mathrm{H}] \mathbf{c} = E^v \mathbf{S} \mathbf{c} ,
\end{equation}

\noindent where the vector $\mathbf{c}$ has components $c_{n,l}^i$, the matrix 
elements of $[\mathrm{H}] $ are given by

\begin{equation}\label{eq:matrix-elements-variational-H}
 [\mathrm{H}]_{n,l,n',l'}^{i,j} = \left\langle B_n P_l \right| \mathrm{H}^{i,j} 
\left| B_{n'} P_{l'} \right\rangle , 
\end{equation}

\noindent and the matrix elements of the overlapping matrix $\mathbf{S}$ can be 
written as

\begin{equation}\label{eq:overlapping-matrix-elements}
[\mathbf{S}]_{n,l,n',l'}^{i,j} = \left\langle B_n P_l | 
 B_{n'} P_{l'} \right\rangle \delta_{i,j},
\end{equation}

\noindent where $\delta$  is the usual Kronecker delta.

If the basis set size is $N$, then the Hamiltonian matrix in 
Equation~\ref{eq:generalized-variational-eigrnvalue-problem} is a $8N\times 8N$ 
matrix, because of this and since the number of $B$-splines functions is 
related to the number of $knots$ the Hamiltonian matrix may become very large. 
Anyway, the relationship in Equation~\ref{eq:b-splines-orthogonality} imposes 
that the Hamiltonian matrices would be  sparse matrices indeed.  

To evaluate the matrix elements in 
Equation~\ref{eq:matrix-elements-variational-H} it is necessary to write down 
all the Cartesian derivatives in the differential operator $\mathrm{H}$ as  
derivatives on the spherical coordinates, then all the integrals in radial 
coordinates can be obtained using a highly efficient Gauss-Legendre 
quadrature method. The integrals that depend on the polar angle can be 
calculated analytically. 

In all the numerical evaluations, the radial coordinate was considered bounded 
for a value $R_{max}$ large enough, in particular quite larger that the 
characteristic radii of the QD (core, internal and external well radius, etc).

For a layered QD with a single well, the integration interval for the radial 
coordinate $[0,R_{max}]$ was split in three subintervals, each one corresponds 
to the core, well and barrier parts of the QD, {\em i.e.} $[0, R_c]$, $[R_c, 
R_1]$ y $[R_1, R_{max}]$. In each interval the distribution of the $knots$ was 
chosen uniform, with $45$ $knots$ for the first and second subintervals, while 
for the last one the number of $knots$ was taken equal to $30$. Besides $ 
R_{max} = 50\,nm$.

For a layered QD with two wells, the interval $[0, R_{max}]$ was split in five 
subintervals $[0, 
R_c]$,
$[R_c, R_1]$, $[R_1, R_2]$, $[R_2, R_3]$ y $[R_3, R_{max}]$. The distribution 
of $knots$ in each subinterval was chosen uniform, with $25$ $knots$ for the 
first, second, third and fourth subintevals, while for the last subinterval 
$20$ $knots$ were used. In this case $R_{max} = 30\,nm$.

\end{document}